\documentclass[preprint,aps,amsmath,superscriptaddress,nofootinbib,tightenlines]{revtex4}
\usepackage{graphicx,psfrag}
\usepackage{bm}
\usepackage{epsfig}

\def\OMIT#1{}

\newcommand{\nn}{\nonumber}

\newcommand{\bn}{{\bar n}}
\newcommand{\bea}{\begin{eqnarray}}
\newcommand{\eea}{\end{eqnarray}}

\newcommand{\bnQ}{\bar n \!\cdot\! Q}
\newcommand{\bnP}{\bar {\cal P}}

\newcommand{\cP}{{\cal P}}

\newcommand{\mup}{M_\Upsilon}
\newcommand{\mpsi}{M_\psi}
\newcommand{\pup}{p_\Upsilon}

\newcommand{\jpsi}{J/\psi}

\begin{document}

%



\title{Color-Octet $\jpsi$ Production in $\Upsilon$ Decay near the Kinematic Limit}
	
\author{Xiaohui Liu}
\affiliation{Department of Physics and Astronomy,
	University of Pittsburgh,
        Pittsburgh, PA 15260\vspace{0.2cm}}

\date{\today\\ \vspace{1cm} }



\begin{abstract}
Recent experiments by the CLEO III detector at CESR indicate that the $\jpsi$
spectrum produced in $\Upsilon$ decay is in conflict with 
Non-Relativistic QCD (NRQCD) calculations. The measured $\jpsi$ momentum
distribution is much softer than predicted by the color-octet mechanisms. The 
expected peak at the kinematic limit is not observed in the data. However it has 
recent been pointed out that  NRQCD 
calculations break down near the kinematic endpoint due to large perturbative 
and non-perturbative corrections. In this paper we combine NRQCD with
soft collinear effective theory  to study the color-octet 
contribution to the $\Upsilon \to \jpsi + X$ decay in this region of phase 
space. We obtain a spectrum that is significantly softened when including the 
correct degrees of freedom in the endpoint region, giving better agreement with 
the data than previous predictions.
\end{abstract}

\maketitle

\newpage

\section{Introduction}
Since the discovery of the $J/\psi$, heavy quarkonium decay and production 
have served as a laboratory for testing both perturbative and
nonperturbative aspects of QCD dynamics. The large mass $m_Q$ of the heavy
quark inside the quarkonium set a high mass scale ($\sim 2m_Q$) at which the QCD approaches 
the asymptotically free regime and thus one could hope to calculate perturbatively.
On the other hand, hadronization process happens at the much smaller mass scale 
of order $m_Q v^2$, where $v$ is the typical velocity of the heavy quarks in 
the quarkonium. For quarkonium, $m_Q v^2$ is numerically of order 
$\Lambda_{\rm QCD}$ so the involved QCD process also contains nonperturbative
effects.

During the past 15 years, the interactions of the non-relativistic heavy quarks inside the 
quarkonium have been
understood to some extent using the framework of Non-Relativistic QCD 
(NRQCD)~\cite{Bodwin:1995prd,Luke:2000prd}. NRQCD is an effective theory 
that reproduces the physics of full QCD by adding local interactions that 
systematically incorporate relativistic corrections through any given order 
in the heavy quark velocity $v$.  This effective theory provides a generalized 
factorization theorem that includes nonperturbative
corrections to the color-singlet model, including color-octet decay mechanisms.
 All infrared divergences can be factored into nonperturbative matrix elements,
 so that infrared safe calculations of inclusive decay rates are 
possible~\cite{Bodwin:1992prd}.
NRQCD solves some important phenomenological problems in quarkonium physics. 
For instance, it provides the most convincing explanation to the 
surplus $\jpsi$ and $\psi'$ production at the Tevatron~\cite{Braaten:1995prl},
in which a gluon fragments into a color-octet $c\bar{c}$ pair in a pointlike
color-octet S-wave state which evolves nonperturbatively into the charmonium
states plus light hadrons. The NRQCD factorization formalism allows these 
fragmentation procedures to be factored into the product of short distance 
coefficients and long distance NRQCD matrix elements among which 
the leading one is  
$\langle {\cal O}^{\bf 8}_{\psi(\psi')}[{}^3S_1]\rangle$ where 
$ {\cal O}^{\bf 8}_{\psi(\psi')}$ are local $4$-fermon operators in terms of 
the fields of NRQCD.

There are, however, some problems that remain to be solved. 
One challenging problem is with the polarization of $\jpsi$ 
at the Tevatron. The same mechanism that produces the $\jpsi$ described above 
predicts the $\jpsi$ should become transversely 
polarized as the transverse momentum $p_\perp$ becomes 
much larger than $2m_c$~\cite{Cho:1995plb}. 
Though the theoretical prediction is consistent with the experimental data at 
intermediate $p_\perp$, at the largest measured values of $p_\perp$ the 
$\jpsi$ is observed to be slightly longitudinally polarized and 
discrepancies at the $3\sigma$ level are seen in both 
prompt $\jpsi$ and $\psi'$ polarization measurements~\cite{Affolder:2000prl}.

Recently, a new problem arose as a result of measurements of the 
spectrum of $\jpsi$ produced in the $\Upsilon(1S)$ decay by the CLEO III 
detector at CESR~\cite{Briere:2004prd}. NRQCD calculations have been made
for the production of $\jpsi$ through both color-singlet and color-octet 
configurations~\cite{Li:2000plb,Cheung:1996prd}. Theoretical calculations
predict that the color-singlet process $\Upsilon(1S)\to \jpsi c\bar{c}g+X$
features a soft momentum spectrum.
Meanwhile, the theoretical estimates based on color-octet contribution 
indicates that the momentum spectrum peaks near the kinematic 
endpoint~\cite{Cheung:1996prd}. In contrast to the theoretical predictoins,
the experimentally measured momentum spectrum is significantly softer than
predicted by the color-octet model and somewhat softer than the color-singlet
case~\cite{Briere:2004prd}.    

The NRQCD predictions break down in the endpoint region because the effective field theory 
does not contain the correct degrees of freedom to describe the physics.  NRQCD contains soft
 quarks and gluons, but it does not contain quarks and gluons moving collinearly.
The correct effective theory to use in situations where there is both soft and collinear physics is 
Soft-Collinear Effective 
Theory (SCET)~\cite{BauerO:2001prd,BauerS:2001prd,Bauer:2001plb,Bauer:2002prd}.  
A combination of SCET and NRQCD has been successful in reproducing the shape of the measured $\jpsi$ momentum spectrum in 
$e^+e^-\to \jpsi +X$~\cite{FlemingIntro:2003prd}. SCET has the power to 
describe the endpoint regime by including the light energetic degrees of 
freedom. In addition, renormalization group equations of SCET can be used to
resum large perturbative logarithmatic correctoins. Nonperturbative martix
will occur naturely in deriving the factorizatoin theorem using SCET.

In this paper, we use SCET to study the color-octet 
contribution to the $\Upsilon \to \jpsi + X$ decay near the endpoint. 
We derive the factorization theorm in SCET for this process.
We find that the spectrum is significantly softened 
when including perturbative up to leading logarithms (LL) 
and nonperturbative corrections near the 
endpoint, giving better agreement with the data than the previous 
predictions.

\section{Factorization and Matching}\label{FactMat}

In this section, we briefly derive the SCET factorization theorem 
for $\Upsilon \to \jpsi + X$ near the endpoint. A more detailed 
derivation will be presented in the Appendix. The derivation is 
similar to radiative $\Upsilon$ decay~\cite{Fleming:2003prd}, 
which we refer to for details. However, for the process we discussed here, 
it involves the decay of a heavy quarkonium into another heavy quarkonium,
thus we should combine SCET with two independent NRQCD's for these two onia 
systmes. The factorization for a similar process 
$B \to \jpsi + X_s$ has been discussed in Ref.~\cite{Bobeth:2008prd}.

Near the endpoint regime, a new factorization formula is required 
since the NRQCD does not
include all the relevant physical degrees of freedom and thus the 
factorization theorem breaks down. 
This can easily be seen when we analyze the kinematics at the endpoint. 
To do so, we work in the centre-of-mass (COM) frame, and introduce 
light cone coordinates. By introducing the parameter 
$x = (E_\psi+ p_\psi)/\mup$, we have
\bea
&&\pup^\mu = \,
\frac{\mup}{2}n^\mu + \frac{\mup}{2}\bn^\mu + k^\mu_\Upsilon\,, \nn \\
&&p_\psi^\mu = \,
\frac{\mpsi^2}{2x\mup}n^\mu + \frac{x\mup}{2}\bn^\mu +k^\mu_\psi\,, \nn \\
&&p_X^\mu = \,
\frac{\mup}{2}\left[\left(1-\frac{r}{x}\right)n^\mu + \,
(1-x)\bn^\mu \right] + k_X^\mu \,.
\eea
Here $n=(1,0,0,1)$ and $\bn = (1,0,0,-1)$, we 
have defined $r = m_c^2/m_b^2$, and we also assumed
that $M_\psi = 2 m_c$ and $\mup = 2m_b$. 
$k_\Upsilon^\mu$ and $k_\psi^\mu$ are the residual momentum 
of the $Q\bar Q$ pair inside the $\Upsilon$ and $\jpsi$ respectively. Near the kinematic 
endpoint, the variable $x \to 1$ and thus the jet invariant mass 
approaches zero.
In NRQCD, an expansion of $k^\mu/m_X$ is performed and hence the 
jet mode is integrated out, which is only
valid when it has a large invariant mass, i.e., away from the endpoint. 
As $x \to 1$, the jet becomes energy is large, but the invariant mass becomes small,
with $k^\mu/m_X$ of order $1$. Hence we must
keep $k^\mu/m_X$ to all orders. As a result, the standard NRQCD 
factorization breaks down at the endpoint.  SCET is the appropriate framework 
for properly including the collinear modes needed in the endpoint  
in order to make reasonable predictions.

\begin{figure}[t]
\begin{center}
\includegraphics[width=8cm]{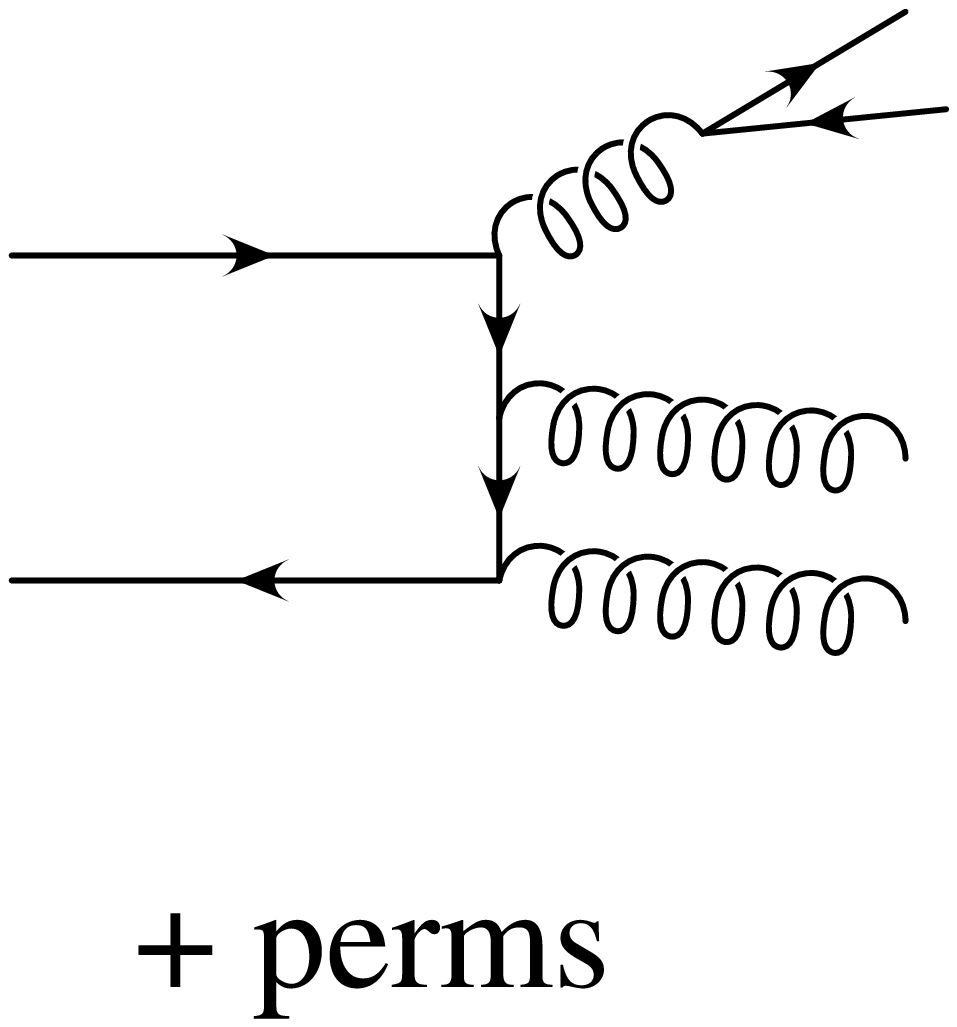}
\caption{\small 
QCD production amplitude for $\Upsilon \to \jpsi + X$. 
The $\jpsi$ is produced  in a color-octet and becomes 
a color-singlet by emitting a soft gluon. 
There is another contribution to this process with only one gluon emitted,
which is suppressed by an order of $\alpha_s$.
\label{match}}
\end{center}
\end{figure}

To derive the factorization theorem in SCET, we start from the optical theorem 
in which the decay rate can be written
as
\bea
2E_\psi \frac{\mathrm{d}\Gamma}{\mathrm{d}^3p_\psi} \,
 = \frac{1}{16\pi^3 \mup}\sum_X \int \mathrm{d}^4y e^{-iq\cdot y}\,
\langle \Upsilon |{\cal O}^\dagger(y)|\jpsi+X \rangle \,
\langle \jpsi + X |{\cal O}(0)|\Upsilon \rangle  \,,
\eea
where the summation includes integration over phase space of $X$. The SCET operator  
${\cal O}$ is 
\bea
{\cal O} = \sum_i \sum_{\omega} e^{-i(M_\Upsilon v+\bnP \frac{n}{2})\cdot y} \,
C_i(\mu,\omega){\cal J}_i(\omega) \,,
\eea
where the Wilson
coefficient $C_i$ is obtained by matching from QCD to SCET at some hard scale 
$\mu = \mu_H$ and the SCET current function ${\cal J}_i(\omega)$ is contrained by 
the gauge invariance. For instance, in our case, to leading order the 
non-vanshing SCET current will be of the form 
\bea\label{OSCET}
{\cal J}(\omega) = \Gamma_{abc}^{\alpha\beta\mu\nu} \,
\left[B_{\alpha, \omega_1}^{a\perp}B_{\beta,\omega_2}^{b\perp} \right]\,
\left[\chi^\dagger_{{\bar b}}\left( \Lambda_1 \cdot \sigma \right)_\nu\,
 \psi_b\right]\,
\left[\chi^\dagger_{{\bar c}} \left(\Lambda_2 \cdot  \sigma \right)_\mu\,
 T^c \psi_c \right]\,.
\eea
Here $\Gamma_{abc}^{\alpha\beta\mu\nu}$ is a hard coefficient containing the 
color and spin structures which is obtained by matching onto the QCD 
Feynman diagrams shown in fig.~\ref{match}. 
The matching gives
\bea
\Gamma_{abc}^{\alpha\beta\mu\nu}\,
= \frac{2ig^4}{N_c}\frac{1}{1-r}\frac{M_\psi}{M_\Upsilon}\,
\frac{1}{M_\psi^2} d_{abc}g_\perp^{\alpha\beta}\,
\left(g_\perp^{\mu\nu}+\bn^\mu n^\nu \right) \,,
\eea
where we have chosen the coefficient so that the Wilson coeffcient 
$C(\mu,\omega)$ is $1$ at the hard scale $\mu_H$. 
The $\Lambda$'s boost the $\jpsi$ or $\Upsilon$ from the COM frame to a frame where 
those quarkonia have arbitrary four-momentum. 
$\psi$ and $\chi$ are the heavy quark and antiquark fields which  
create or annihilate the constituent heavy (anti-)quarks inside the quarkonia.
The collinear gauge invariant field strength is built out of the collinear 
gaugle field $A^\mu_{n,q}$ 
\bea  
B_\perp^\mu = \frac{-i}{g_s}W_n^\dagger \left( \, 
{\cal P}_\perp^\mu + g_s(A^\mu_{n,q})_\perp \right)W_n \,,
\eea
where
\bea
W_n = \sum_{\rm perms} \exp \left( -g_s \frac{1}{\bnP} \bn \cdot A_{n,q} \,
\right) \,,
\eea
is the collinear Wilson line. The operator $\cP$ picks out the large momentum
label~\cite{Bauer:2001plb}.

We can decouple the usoft modes from the collinear degrees of 
freedom by making the field redefinition~\cite{Bauer:2002prd}
\bea\label{Atrans}
A_{n,q}^\mu \to YA_{n,q}^\mu Y^\dagger \,,
\eea
where $Y$ is the usoft Wilson line made out of the usoft gauge fields. In such
a way, we can separate the collinear physics from the usoft and obtain the 
factorization theorem in SCET. Meanwhile using similar arguments presented
in Ref.~\cite{Bobeth:2008prd} allows us to factorize the matrix element into
the convolution of the shape functions for $\Upsilon$ in color-singlet 
configuration and $\jpsi$ in color-octet one.
Following this procedure, the decay rates can be written as 
a convolution of soft shape functions and the jet function with an 
overall hard coefficient. By introducing $z=E_\psi/m_b$, 
we get the decay rate of the form
\bea\label{Facform}
\frac{\mathrm{d}\Gamma}{\mathrm{d}z} = \Gamma_0 P[z,r] \,
\sum_\omega |C(\omega,\mu)|^2\,
\int \mathrm{d}k^+ \int \mathrm{d}l^+ J_\omega(k^+)  \,
S_\psi(l^+)  S_\Upsilon(\mup(1-x)-k^+-l^+) \,,
\eea
where $P[z,r]= 8\pi \sqrt{z^2-4r}/(1-r)$ is a kinematic factor and
%
\bea
\Gamma_0  = \frac{\pi \alpha_s^4}{18} \frac{N_c^2 -4}{N_c^3} \,
\frac{2+r}{1-r} \frac{1}{m_b^2 m_c^3} \,
\langle \Upsilon | {\cal O}^{\bf 1}_\Upsilon[{}^3S_1] | \Upsilon \rangle \,
\langle {\cal O}^{\bf 8}_\psi[{}^3S_1] \rangle \,. 
\eea
We have used spin symmetry~\cite{Braaten:1996prd} 
\bea
&&\Lambda_i^{\delta}\Lambda_j^{\delta'} \langle \dots {\bf \sigma}^i \dots \,
{\bf \sigma}^j \dots \rangle = \nn \\
&&\hspace{3.ex} \frac{1}{3}\delta^{ij}\Lambda_i^{\delta}\Lambda_j^{\delta'} \,
\langle \dots {\bf \sigma}^k \dots \,
{\bf \sigma}^k \dots \rangle \,,
\eea
to simplify the matrix elements, and applied the
identity $\delta^{ij}\Lambda_i^{\delta}\Lambda_j^{\delta'} 
= (v^{\delta}v^{\delta'}-g^{\delta \delta'})$, where $v^\delta$ is the four-velocity
of the $\Upsilon$ or $\jpsi$.

The shape function for $\jpsi$ is defined as
\bea\label{PsiShape}
S_{\psi}(l^+) = \int \frac{\mathrm{d}y^-}{4\pi}\,
e^{-\frac{i}{2}l^+y^-}\,
\frac{\langle 0|\left[\,
\chi_{\bar c}^\dagger \sigma_i Y\tilde{Y} T^k \tilde{Y}^\dagger Y^\dagger\,
\psi_c(y^-) \,
a^\dagger_\psi a_\psi \,
\psi_c^\dagger \sigma_i Y\tilde{Y} T^k \tilde{Y}^\dagger Y^\dagger\,
\chi_{\bar c} \,
\right]|0\rangle}{4m_c\langle {\cal O}^{\bf 8}_\psi[{}^3S_1]\rangle}\,,
\eea
where we have made the field redefinition in Eq.~(\ref{Atrans}) for the two 
collinear gluons in the final state by introducing two different usoft Wilson lines
$Y$ and $\tilde{Y}$.
For $\Upsilon$ we have the shape function
\bea
S_{\Upsilon}(l^+) =  \int \frac{\mathrm{d}y^-}{4\pi}\,
e^{-\frac{i}{2}l^+y^-}\,
 \frac{\langle\Upsilon|\chi_{\bar b}^\dagger \sigma_i \,
\psi_b(y^-)\psi_b^\dagger \sigma_i \chi_{\bar b}|\Upsilon\rangle}{4m_b\langle\Upsilon| {\cal O}^{\bf 1}_\Upsilon[{}^3S_1]|\Upsilon\rangle}
\eea
respectively. Both shape functions are normalized so that 
$\int \mathrm{d}l^+S_{\psi,\Upsilon}(l^+) = 1$.

\begin{figure}[t]
\begin{center}
\includegraphics[width=8cm]{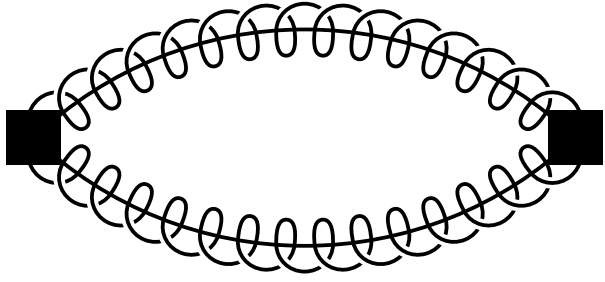}
\caption{\small 
Feynman diagram for the leading-order jet function. The spring with a line through
it represents a collinear gluon.
\label{vacloop}}
\end{center}
\end{figure}

The jet function is given by 
\bea
&&\langle 0| \left[ B_{\alpha}^{a\perp}B_{\beta}^{b\perp}(y)\,
 B_{\alpha'}^{a'\perp}B_{\beta'}^{b'\perp}(0)\right] |0 \rangle \nn\\
&=& \frac{i}{2}\,
(g_{\alpha\alpha'}g_{\beta\beta'}\delta^{aa'}\delta^{bb'} \,
+ g_{\alpha\beta'}g_{\beta\alpha'}\delta^{ab'}\delta^{ba'})\,
\delta_{\omega\omega'}\,
\int \frac{\mathrm{d}k^+}{2\pi} \delta^{(2)}(y^\perp )\,
\delta(y^+) e^{-\frac{i}{2}k^+y^-} J_\omega(k^+) \,.
\eea
To the leading order, the jet function can be calculated by evaluating the
diagram shown in Fig.~\ref{vacloop} , which gives
\bea
J_\omega(k^+)=\frac{1}{8\pi}\Theta(k^+)\int_0^1 \mathrm{d}\xi\,
 \delta_{\xi,(\bnQ+\omega)/(2\bnQ)}\,,
\eea
where $Q$ is the total four momentum carried by the jets.

The $\jpsi$ shape function can formally be written as
\bea
S_\psi(l^+) = \frac{\langle 0|\left[\,
\chi_{\bar c}^\dagger \sigma_i Y\tilde{Y} T^k \tilde{Y}^\dagger Y^\dagger\,
\psi_c \,
\delta(i n\cdot \partial - l^+)
a^\dagger_\psi a_\psi \,
\psi_c^\dagger \sigma_i Y\tilde{Y} T^k \tilde{Y}^\dagger Y^\dagger\,
\chi_{\bar c} \,
\right]|0\rangle}{4m_c\langle {\cal O}^{\bf 8}_\psi[{}^3S_1]\rangle}\,,
\eea
and to lowest order in $v^2$, $S_{\Upsilon}(l^+) \to \delta(l^+)$. By 
integrating over $k^+$ and $l^+$ in Eq.~(\ref{Facform}), we find
the tree level decay rates become
\bea
\frac{\mathrm{d}\Gamma}{\mathrm{d}z} = \Gamma_0 \tilde{P}[z,r] \Theta (1-x) \,,
\eea
with $\tilde{P}[z,r] = P[z,r]/8\pi$.  This can easily be seen to reproduce the
tree level calculation of NRQCD~\cite{Cheung:1996prd}.

\section{Running}

Effective field theories provide a powerful tool to sum logarithms
by using the renormalization group equations (RGEs). Logarithms of the 
ratio of different scales arise naturally in perturbation theory, which 
can cause a breakdown of the perturbative expansion when those scales are 
well separated.  By matching onto an effective theory, the large scale is removed to
be replaced by a running scale $\mu$ and the effective
operators are run from a high scale to the low scale using the RGEs, which  
sum all large logarithms of the ratio of scales into an overall factor. 

In our case, there are logarithms of the form $\log(1-x)$ that will appear in 
the perturbation series. Near the endpoint, $x \to 1$, these 
become large, and must be resummed.  
In this section, we will apply the RGEs of SCET to sum these large logarithms.

In the previous section, we have matched QCD onto the SCET operator by
intergrating out the hard scale $\mu_H$, replacing it with a running scale 
$\mu$. We now run the operator from this hard scale to the collinear scale. To do 
so, we calculate the counterterm for the operator to determine the anomalous
 dimension, and then use this in the RGEs. 

\begin{figure}[t]
\begin{center}
\includegraphics[width=12cm]{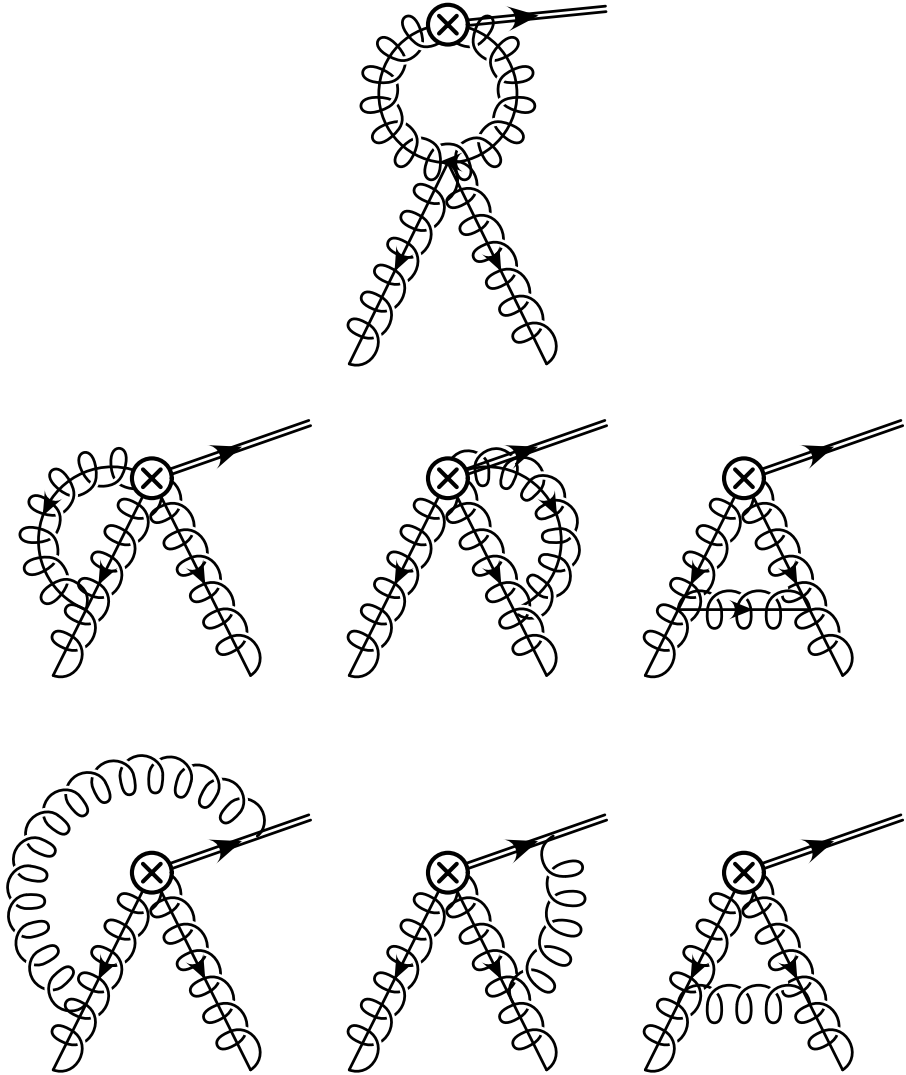}
\caption{\small 
One-loop order diagrams needed to calculate the counterterm to the color-octet
operator.The double line presents the $\jpsi$ fields in color-octet 
configuration while the spring lines are the soft gluons.  
\label{HarCorr}}
\end{center}
\end{figure} 

The one-loop corrections to the SCET operator in Eq.~(\ref{OSCET}) is given by 
the graphs in Fig.~\ref{HarCorr}. Evaluating these diagrams gives the 
divergent term
\bea
{\cal A}_{1-\rm{loop}} &=& \sum_{\omega} \frac{\alpha_s C_A}{4\pi}\,
\left\{ \left[\frac{1}{\epsilon^2}+\frac{1}{\epsilon}\,
\left(2 + \log \frac{\mu^2}{\bnQ^2/r}\right) \right] \right. \nn \\
&& \qquad + \,
\left. \frac{1}{\epsilon}\,
\left[ \frac{\omega(\bnQ+\omega)}{\bnQ(\bnQ-\omega)}\,
\log \frac{\bnQ+\omega}{2\bnQ} \,
-  \frac{\omega(\bnQ-\omega)}{\bnQ(\bnQ+\omega)}\,
\log \frac{\bnQ-\omega}{2\bnQ}\right] \right\} \times {\cal A}_0 \,. \nn \\
\eea
The calculation lets us estimate the hard scale be $\mu_H = \bnQ/\sqrt{r}$ 
which will minimize
the logarithm. The divergent piece must be canceled by 
$Z_{\bf 8}Z_3/Z_{\cal O} -1$, where $Z_{\cal O}$ is the couterterm for the 
operator in SCET, $Z_3$ is the gluon wave function counterterm
\bea
Z_3 = 1 + \frac{\alpha_s}{4\pi}\frac{1}{\epsilon}\,
\left(\frac{5}{3}C_A - \frac{4n_F}{3}T_F\right) \,,
\eea
and $Z_{\bf 8}$ is the counterterm of color-octet $\jpsi$ operator
\bea
Z_{\bf 8} = 1 + \frac{\alpha_s C_A}{4\pi\epsilon}\,.
\eea
This leads to 
\bea\label{ZO}
Z_{\cal O} -1 &=& \sum_{\omega} \frac{\alpha_s C_A}{4\pi}\,
\left\{ \left[\frac{1}{\epsilon^2}+\frac{1}{\epsilon}\,
\left(\log \frac{\mu^2}{\bnQ^2/r}\right)\,
+\frac{1}{\epsilon} \left(\frac{14}{3}-\frac{4n_F}{3}\frac{T_F}{C_A} \right)\, 
\right] \right. \nn \\
&& \qquad + \, 
\left. \frac{1}{\epsilon}\,
\left(
+ \frac{\omega(\bnQ+\omega)}{\bnQ(\bnQ-\omega)}\,
\log \frac{\bnQ+\omega}{2\bnQ} \,
-  \frac{\omega(\bnQ-\omega)}{\bnQ(\bnQ+\omega)}\,
\log \frac{\bnQ-\omega}{2\bnQ}\right) \right\} \,. \nn \\
\eea
From Eq.~(\ref{ZO}), we can extract the anomalous dimensioin of the operator
through the standard method. Using the anomalous dimension in the RGE for the 
color-octet Wilson coefficient and running from the hard scale down to the 
collinear scale gives
\bea
|C(\xi,\mu_c)|^2 = \left[\frac{\mu_c^2}{\bnQ^2/r}\right]^{-\frac{2C_A}{b_0}}\,
\left[\frac{\alpha_s(\mu_c^2)}{\alpha_s(\bnQ^2/r)}\right]^{-\frac{8\pi C_A}{\alpha_s(\bnQ^2/r)b_0^2}} \,
\left[\frac{\alpha_s(\mu_c^2)}{\alpha_s(\bnQ^2/r)}\right]^{4\eta[\xi]}\,,
\eea
where
\bea
\eta[\xi] = \frac{C_A}{2b_0}\left[ \left(\frac{14}{3} \,
- \frac{4n_F}{3}\frac{T_F}{C_A} \right) \,
-(2\xi -1)\left(\frac{1-\xi}{\xi}\log (1-\xi)\,
- \frac{\xi}{1-\xi}\log \xi \right) \right]\,,
\eea
with $\xi = (\bnQ+\omega)/(2\bnQ)$, $b_0 = 11C_A/3-2n_F/3$, and the collinear scale 
$\mu_c^2 \approx m_X^2$. 

At the collinear scale, the jet mode can be regarded as large and can be 
integrated out. The decay rate can be further run down to the soft scale
$\mu_s$. To do this, we first note that the decay rate can be modified to
\bea
\frac{\mathrm{d}\Gamma}{\mathrm{d}z} \,
= \Gamma_0 \tilde{P}[z,r]\int_0^1 \mathrm{d}\xi |C(\xi,\mu_c)|^2 \,
\int \mathrm{d}l^+ \mathrm{d}l^{+'} \Theta(M_\Upsilon(1-x)-l^+)\,
{\cal U}_s(l^+ - l^{+'},\mu_c,\mu_s) S_\psi(l^{+'},\mu_s)\,, \nn \\
\eea
since in Ref.~\cite{Rothstein: 1997plb}, it was shown that
\bea
\langle\Upsilon|\chi_{\bar b}^\dagger \sigma_i \,
\psi_b \,
\Theta(i n\cdot \partial+ \mup(1-x)-l^+ )\,
\psi_b^\dagger \sigma_i \chi_{\bar b}|\Upsilon\rangle
&=& \nn \\
&& \hspace{-12.ex}\,
\Theta(\mup(1-x)-l^+) \langle\Upsilon|\chi_{\bar b}^\dagger \sigma_i \,
\psi_b \,
\psi_b^\dagger \sigma_i \chi_{\bar b}|\Upsilon\rangle \,.
\eea
And here we introduced an evolution kernel ${\cal U}_s$ as in 
Ref.~\cite{Fleming:2007prd}. 
The soft shape function has evolution through ${\cal U}_s$ which will sum the 
large logarithms between $\mu_s$ and $\mu_c$. 

\begin{figure}[t]
\begin{center}
\includegraphics[width=12cm]{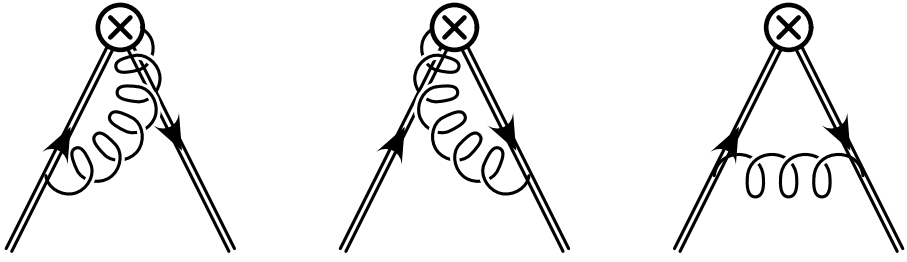}
\caption{\small 
One loop corrections to the $\jpsi$ soft shape function defined in 
Eq.~(\ref{PsiShape}).
\label{LoopSoft}}
\end{center}
\end{figure} 

The evolution kernel can be calculated explicitly~\cite{Fleming:2007prd}, once
we figure out the anomalous dimension for the soft shape function. To one-loop
order, calculating the diagrams in fig.~\ref{LoopSoft}, we get
\bea
Z_{S_{\psi}}-1 = \frac{\alpha_s}{2\pi}C_A \,
\left[\left(-\frac{1}{\epsilon^2}\,
-\frac{1}{\epsilon}\log \frac{\mu^2}{r\mup^2} + \frac{1}{\epsilon}\right)\,
\delta(k^+) + \frac{2}{\epsilon}\frac{1}{\mup}\left( \,
\frac{\mup \Theta(k^+)}{k^+} \right)_+ \right]\,.
\eea
Therefore we ${\cal U}_s$ is
\bea
\label{Us}
{\cal U}_s(l^+ - l^{+'}) =\,
 \frac{e^{\tilde{K}}(e^{\gamma_E})^{\tilde{\omega}}}{\mu_s \Gamma(-\tilde{\omega})}\,
\left[ \frac{\mu_s^{1+\tilde{\omega}} \Theta(l^+-l^{+'})}{(l^+ - l^{+'})^{1+\tilde{\omega}}}\right]_+ \,,
\eea
with $\tilde{\omega}$ is defined as
\bea
\tilde{\omega} = -\frac{2C_A}{\pi}\int_{\alpha_s(\mu_s)}^{\alpha_s(\mu_c)}\,
\frac{\alpha \mathrm{d}\alpha}{\beta[\alpha]}\,
 = \frac{4C_A}{b_0}\log\frac{\alpha_s(\mu_c)}{\alpha_s(\mu_s)}\,,
\eea
where  $\beta[\alpha_s] = -(11C_A/3 - 2n_F/3)\alpha_s^2/(2\pi)$.  Note that $\tilde{\omega} <0$.  We defined a function 
$\tilde{K}_\gamma$
\bea
\tilde{K}_\gamma = \frac{C_A}{\pi}\int_{\alpha_s(\mu_s)}^{\alpha_s(\mu_c)}
\frac{\alpha \mathrm{d}\alpha}{\beta[\alpha]}\left(1+\log r\right)\,
=-\frac{2C_A}{b_0}(1+\log r) \log \frac{\alpha_s(\mu_c)}{\alpha_s(\mu_s)}\,,
\eea
which is related to $\tilde{K}$ in Eq.~(\ref{Us}) by 
\bea
\tilde{K} &=& \tilde{K}_\gamma - \frac{2C_A}{\pi}\,
\int_{\alpha_s(\mu_s)}^{\alpha_s(\mu_c)}\,
\frac{\alpha\mathrm{d}\alpha}{\beta[\alpha]} \,
\int_{\alpha_s(\mu_s)}^\alpha \frac{\mathrm{d}\alpha'}{\beta[\alpha']}\nn\\
&=& \tilde{K}_\gamma \,
+ \frac{8\pi C_A}{b_0^2 \alpha_s(\mu_c)}\left( \,
\frac{\alpha_s(\mu_c)}{\alpha_s(\mu_s)}-1 - \,
\frac{\alpha_s(\mu_c)}{\alpha_s(\mu_s)}\,
\log \frac{\alpha_s(\mu_c)}{\alpha_s(\mu_s)} \right) \,.
\eea
The soft scale is $\mu_s^2 \sim r M_\Upsilon^2(1-x)^2$.

Gathering all the pieces we have, we find the resumed decay rate
\bea\label{DRates}
\frac{\mathrm{d}\Gamma}{\mathrm{d}z} =&& \,
\Gamma_0 \tilde{P}[z,r] \,
 \left[\frac{\mu_c^2}{\bnQ^2/r}\right]^{-\frac{2C_A}{b_0}}\,
\left[\frac{\alpha_s(\mu_c^2)}{\alpha_s(\bnQ^2/r)}\right]^{-\frac{8\pi C_A}{\alpha_s(\bnQ^2/r)b_0^2}} \,
\int_0^1\mathrm{d}\xi 
\left[\frac{\alpha_s(\mu_c^2)}{\alpha_s(\bnQ^2/r)}\right]^{4\eta[\xi]} \nn \\
&&\times \frac{e^{\tilde{K}}(e^{\gamma_E})^{\tilde{\omega}}}{\mu_s \Gamma(-\tilde{\omega})}\,
\int\mathrm{d}l^+ \mathrm{d}l^{+'}\Theta(\mup(1-x)-l^+) 
\left[ \frac{\mu_s^{1+\tilde{\omega}} \Theta(l^+-l^{+'})}{(l^+ - l^{+'})^{1+\tilde{\omega}}}\right]_+ S_\psi(l^{+'},\mu_s)  \,.
\eea

\section{Phenomenology}
The decay rate from the previous section, Eq.~(\ref{DRates}), summed up the 
leading logarithmic corrections which are important near the kinematic 
endpoint. Away from that region, the
logarithms that we have summed are not important and contributions that we 
neglected in the endpoint become important. We therefore would like to 
interpolate between the leading order color-octet contribution in NRQCD away 
from the endpoint and the resummed result near the endpoint. We thus 
define the interpolated differential rate as
\bea
\frac{\mathrm{d}\Gamma}{\mathrm{d}y} = (1-y)\left(\,
\frac{\mathrm{d}\Gamma}{\mathrm{d}y}\right)_{\rm NRQCD} \,
+ y \left(\,
\frac{\mathrm{d}\Gamma}{\mathrm{d}y}\right)_{\rm SCET} \,.
\eea
Here, in order to compare with the data, we have used the scaled momentum 
defined as $y = p_\psi/p_\psi^{max}$. We see that as $y \to 1$ the first term 
vanishes, leaving only the SCET contribution in the endpoint region.

To proceed, we need the soft shape fucntion of $\jpsi$ that appears in 
Eq.~(\ref{DRates}). We will apply a modified version of a model used in the 
decay of B mesons~\cite{Leibovich:2002plb},
\bea\label{Sfun}
f(\hat{l}^+) = \frac{1}{\Lambda}\frac{a^{ab}}{\Gamma(ab)}(\eta-1)^{ab-1}\,
e^{-a(\eta-1)} \Theta(\eta-1)\,,
\eea
with $\eta = \hat{l}^+/\Lambda$. Here $\Lambda = M_\psi - M$ is of order
$\Lambda_{\rm QCD}$, and $a$ and $b$ are adjustable parameters of order $1$. 
In our case, we choose $a=1$ and $b=2$. $\Lambda$ was determined so that
the first and the second moments of the shape function
\bea
m_1&=&\int_\Lambda^\infty \mathrm{d}\hat{l}^+ \,
f(\hat{l}^+) = \Lambda (b+1)\,, \nn\\
m_2&=&\int_\Lambda^\infty \mathrm{d}{\hat{l}}^+(\hat{l}^+ )^2\,
 f(\hat{l}^+) = \Lambda^2\,
\left(\frac{b}{a}+(b+1)^2\right)\,, 
\eea
take the value $890{\rm MeV}$ and $(985{\rm MeV})^2$ respectively. 

\begin{figure}[t]
\begin{center}
\includegraphics[width=12cm]{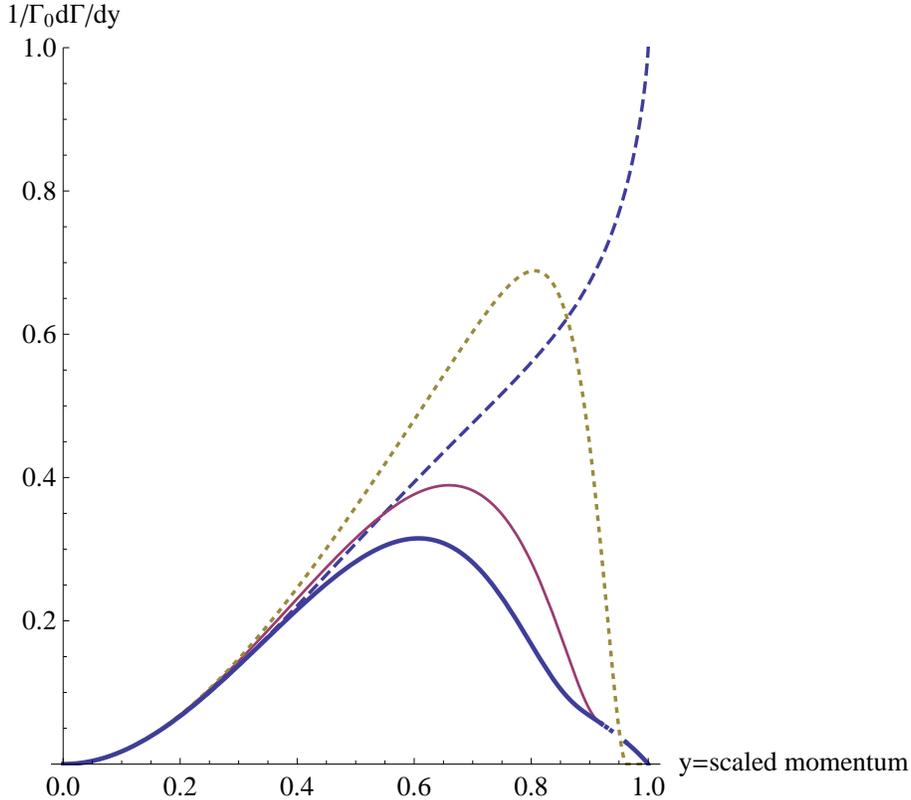}
\caption{\small 
Comparison between the NRQCD results and SCET predictions 
normalized to the NRQCD decay rate at the end-point.
Here $y= p_\psi/p^{max}_\psi$ is the scaled momentum.  The short 
dashed line is the NRQCD 
decay rate only and the dotted line is the NRQCD decay rate convoluted with 
the shape function. The solid 
thin line includes only the perturbative resummed 
interpolated decay rates without convoluted with the soft shape function.
The solid thick line presents the interpolated decay rates 
convoluted with the shape function.
\label{plott1}}
\end{center}
\end{figure}

We show the results of resumming in fig.~\ref{plott1}.  The short 
dashed line is the 
NRQCD decay rate only and the dotted line is the NRQCD decay rate convoluted 
with the shape function. The thin line includes only the perturbative resummed 
interpolated decay rates without convoluted with the soft shape function. 
 The solid thick line presenting our final result is 
the interpolated decay rate
convoluted with the shape function in Eq.~(\ref{Sfun}).
 As can be 
seen, the shape function and the perturbative resummation both result in a softer spectrum.  
The combination of the two is softer still.

\begin{figure}[t]
\begin{center}
\includegraphics[width=12cm]{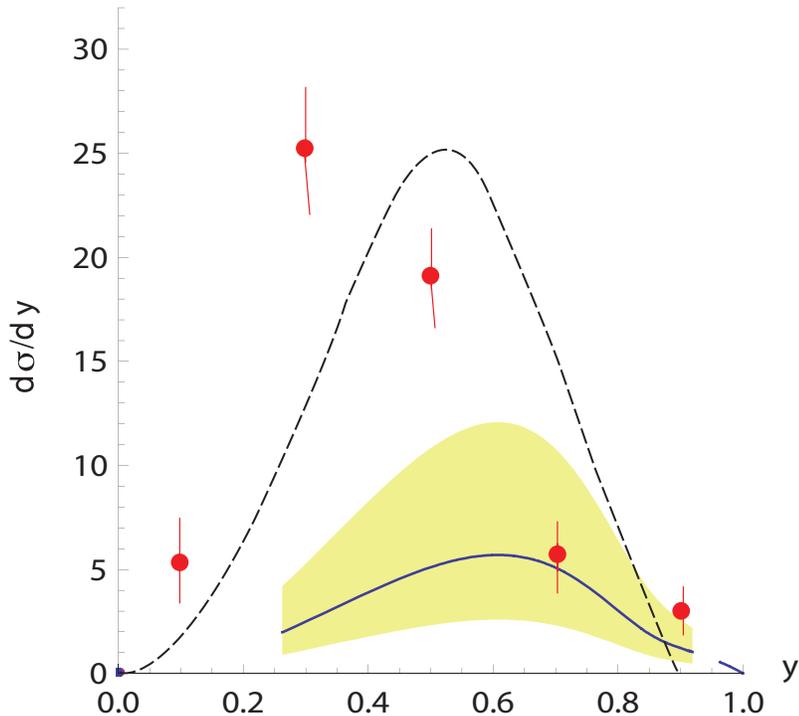}
\caption{\small 
Comparison of  the color-octet contribution to the differential rate to 
the data from
CLEO~\cite{Briere:2004prd}.
The solid thick line presents the interpolated decay rates 
convoluted with the shape function with a choice of 
$\langle {\cal O}^{\bf 8}_\psi[{}^3S_1]\rangle=6.6\times 10^{-3}{\rm GeV}^3$~\cite{Cho:1996prd}.
The shaded band is obtaiend by varying 
$\langle {\cal O}^{\bf 8}_\psi[{}^3S_1]\rangle$ from $0.003{\rm GeV}^3$ to
$0.014{\rm GeV}^3$~\cite{Fleming:1997prd}. Here, we also show the 
color-singlet contribution in long dashed line~\cite{Briere:2004prd,Li:2000plb}. The complete spectrum will involve a combination of both 
the color-octet and color-singlet contributions.
\label{plott2}}
\end{center}
\end{figure}

In fig.~\ref{plott2}, we compare our results with the experimental data from 
CLEO~\cite{Briere:2004prd}. We use the values $m_c = 1.5{\rm\ GeV}$, $m_b = 4.9{\rm\ GeV}$, and 
$\Lambda_{\rm QCD}=0.2{\rm\ GeV}$ so that $\alpha_s(2m_b) = 0.1793$. The solid line represents the color-octet interpolated decay rate convoluted with the shape function.  The 
NRQCD matrix element was chosen to be 
$\langle\Upsilon| {\cal O}^{\bf 1}_\Upsilon[{}^1S_0]|\Upsilon\rangle =2.3{\rm GeV}^3 $. For comparison, we have used in the plot the same value for the 
overall strong coupling evaluated at the scale $2m_c$ as in 
Ref.~\cite{Cheung:1996prd}. The shaded band is obtained by varying the 
NRQCD color-octet matrix
element from 0.003 GeV$^3$ to 0.014 GeV$^3$. Since the numerical value of the 
matrix element 
$\langle {\cal O}^{\bf 8}_\psi[{}^3S_1]\rangle$ is fixed by experimental data,
it has large uncertainties coming both from experiments and theoretical higher order
corrections.  For comparison, we also show the color-singlet contribution as the dashed line~\cite{Briere:2004prd,Li:2000plb}.
 The complete spectrum involves a combination of the color-octet contribution 
we calculated here, and the color-singlet 
component~\cite{Briere:2004prd,Li:2000plb} shown in the figure.


The differential rate predicted by the 
color-octet model is peaked near the end-point region. When convoluted with 
the shape function, the momentum distribution is shifted to the left 
but still close to the kinematic limit. Once we resum the large leading 
logarithms under the framework of SCET, interpolate the result with the NRQCD 
prediction and then convolute with the soft shape function, we find that 
spectrum is significantly softened near the endpoint and the peak is pushed 
further to the left, in better agreement with the data.  We
note here that if we use a high scale for the overall coupling constant, the
color-octet contribution will be much smaller. 
In order to make a consistent comparison of theory to data, one needs 
to treat the endpoint of the color-singlet contribution in SCET and NRQCD, 
which we have not done here and will leave for future work.  In the
perturbative expansion, the color-singlet process is suppressed relative to 
the color-octet one by an factor of $\alpha_s$.   However, there is a large enhancement
due to kinematic factors in the diagram, which can be as large as $360$ that provides a huge 
enhancement to compensate the perturbative suppression~\cite{Li:2000plb}.

\section{Conclusion}

In this paper we studied the color-octet contribution to 
$\Upsilon \to \jpsi+X$ near the kinematic endpoint.  In this regime, the usual 
NRQCD factorization formalism breaks down due to large perturbative and 
nonperturbative corrections. We combined the usual NRQCD effective theory with 
SCET, which contains the correct physical degrees of freedom, to derive a factorizatioin 
theorem for the differential decay rate, $\mathrm{d}\Gamma /\mathrm{d}y$, 
valid in the endpoint region. 
This also allows us to resum large logarithms which appear in the endpoint by 
running the rate from the hard scale to the collinear scale and then to the 
soft scale using the RGE of the effective theory. 
At the soft scale, we are left with NRQCD shape functions.  
Using models for the color-octet shape function, and interpolating away from the endpoint to the leading order NRQCD prediction, we are
able to make predictions about  the color-octet contribution to the decay rate 
over the entire kinematic region.

Though a quantitative comparison to data can be made only when including both 
the color-singlet and color-octet terms, some qulitative conclusions can still
be drawn here. We note that once we sum the large leading logarithms 
using the 
framework of SCET and 
convolute with the soft shape function, the spectrum is significantly 
softened near the endpoint and the peak is broadened and shifted to the left.  
This effect greatly improves the agreement between the data constraints and 
the theoretical predictions.  

We note here that the hard scale we have chosen in our final results is $2m_c$
the same as in ~\cite{Cheung:1996prd}. 
This scale is much smaller than what we expected from our 
one loop order corrections. If we used the hard scale $\mu_H$ decided by 
our caculations, the color-octet contribution will be further suppressed. 
As a result, we expected that the dominant contribution for this process
will be from color-singlet rate. Once including higher order effects as well as
the feed-down of $\psi(2S)$ and $\chi_{cJ}$ to $\jpsi$~\cite{Briere:2004prd}, 
the color-singlet decay rate should also be softened. We expected that 
combining color-singlet and octet contributions will give a good fit to the 
data.


\section{Acknowledgements}

I am grateful to Professor A.~K.~Leibovich for guidances and carefully 
reading the manuscript and checking all the calculations. XL was supported 
in part by the National Science Foundation under Grant No. PHY-0546143.

\appendix\label{Appendix}
\section{Deriving The Factorization Theorem for the Endpoint}\label{factproof}

In this Appendix we show the factorization theorem for 
$\Upsilon \to \jpsi + X$ near the endpoint. We begin with the differential 
decay rate
\bea\label{Optical}
2E_\psi \frac{\mathrm{d}\Gamma}{\mathrm{d}^3p_\psi} \,
 = \frac{1}{16\pi^3 \mup}\sum_X \int \mathrm{d}^4y e^{-iq\cdot y}\,
\langle \Upsilon |{\cal O}^\dagger(y)|\jpsi+X \rangle \,
\langle \jpsi + X |{\cal O}(0)|\Upsilon \rangle  \,,
\eea
where $X$ includes both the usoft $X_u$ sector and the collinear $X_c$ sector. 
The operator ${\cal O}$ is the SCET operator defined 
in Section~\ref{FactMat} of the form
\bea
{\cal O} = e^{-i(\mup v+ \bnP \frac{n}{2})\cdot y}\,
\Gamma^{\alpha\beta\mu\nu}_{abc} \,
{\cal J}^{ab}_{\alpha\beta} {\cal O}^\Upsilon_\nu[{\bf 1}^3S_1]\,
 {\cal O}^{c\psi}_{\mu}[{\bf 8}^3S_1] \,,
\eea
where
\bea
 {\cal J}^{ab}_{\alpha\beta} &=& \,
B_{\alpha, \omega_1}^{a\perp}B_{\beta,\omega_2}^{b\perp} \,, \\
  {\cal O}^\Upsilon_\nu[{\bf 1}^3S_1] &=&\,
 \psi^\dagger_{{b}}\left( \Lambda_1 \cdot \sigma \right)_\nu\,
 \chi_{\bar b} \,, \\
  {\cal O}^{c\psi}_{\mu}[{\bf 8}^3S_1] &=&\,
\psi^\dagger_{{c}} \left(\Lambda_2 \cdot  \sigma \right)_\mu\,
 T^c \chi_{\bar c} \,.
\eea

Inserting the operator into  Eq.~(\ref{Optical}), the ${\cal O}^\dagger (y)$ picks
up an additional phase and the differential rate becomes
\bea\label{Adefin}
2E_\psi \frac{\mathrm{d}\Gamma}{\mathrm{d}^3p_\psi}&=& \,
  \frac{1}{16\pi^3 \mup}\sum_X \int \mathrm{d}^4y \,
e^{-i\mup /2 (1-x)\cdot \bn \cdot y}\,
{\Gamma^{\alpha\beta\mu\nu}_{abc}}^\dagger \,
\Gamma^{\alpha'\beta'\mu'\nu'}_{a'b'c'}\nn\\
&&\hspace{-4ex}\times \langle \Upsilon |\,
{\cal J}^{ab\dagger}_{\alpha\beta}\,
 {\cal O}^{\Upsilon\dagger}_\nu[{\bf 1}^3S_1]\,
 {\cal O}^{c\psi\dagger}_{\mu}[{\bf 8}^3S_1](y) |\jpsi+X \rangle \nn \\
&&\hspace{-2.5ex} \langle \jpsi + X | \,
{\cal J}^{a'b'}_{\alpha'\beta'} {\cal O}^\Upsilon_{\nu'}[{\bf 1}^3S_1]\,
 {\cal O}^{c'\psi}_{\mu'}[{\bf 8}^3S_1](0)|\Upsilon \rangle  \nn \\
&\equiv &  {\Gamma^{\alpha\beta\mu\nu}_{abc}}^\dagger \,
\Gamma^{\alpha'\beta'\mu'\nu'}_{a'b'c'} \,
{\cal A}^{abc,a'b'c'}_{\alpha\beta\mu\nu,\alpha'\beta'\mu'\nu'} \,.
\eea
In the exponent of Eq.~(\ref{Adefin}), we have used 
$q^\mu - \mup v^\mu + \bnP n^\mu/2 \approx \mup/2(1-x)\bn^\mu $.
As mentioned in Section~\ref{FactMat}, we can decouple the usoft modes from 
the collinear degrees of 
freedom using the field redefinition~\cite{Bauer:2002prd}
\bea
A_{n,q}^\mu \to YA_{n,q}^\mu Y^\dagger  \,,
\eea
which modifies $ {\cal O}^{c\psi}_{\mu}[{\bf 8}^3S_1]$ to
\bea
 {\cal O}^{c\psi}_{\mu} \to \,
 Y\tilde{Y}{\cal O}^{c\psi}_{\mu} \tilde{Y}^\dagger Y^\dagger \equiv \,
\tilde{\cal O}^{c\psi}_{\mu} \,.
\eea
Using this field redefinition, we can write 
\bea\label{SAt}
{\cal A}^{abc,a'b'c'}_{\alpha\beta\mu\nu,\alpha'\beta'\mu'\nu'} \,
&=& \frac{1}{16\pi^3 \mup} \,
\int \mathrm{d}^4y \,e^{-i\mup/2(1-x)\bn \cdot y} \nn \\
&&\hspace{-4.ex} \times  \,
\langle \Upsilon |\,
 {\cal O}^{\Upsilon\dagger}_\nu[{\bf 1}^3S_1]\,
 \tilde{{\cal O}}^{c\psi\dagger}_{\mu}[{\bf 8}^3S_1](y) \,
a^\dagger_\psi a_\psi \,
 {\cal O}^{\Upsilon}_{\nu'}[{\bf 1}^3S_1]\,
 \tilde{{\cal O}}^{c'\psi}_{\mu'}[{\bf 8}^3S_1](0)\,
| \Upsilon \rangle \nn \\
&&\hspace{-4.ex} \times  \,
\langle 0 |\,
{\cal J}^{ab\dagger}_{\alpha\beta}(y)
{\cal J}^{a'b'}_{\alpha'\beta'}(0) | 0\rangle  \,,
\eea
where we have introduced an interpolating field,
$a_\psi$, for the $\jpsi$ and used the completeness of states in the usoft 
and collinear sectors
\bea
\sum_{X_u} | J/\psi+X_u \rangle \langle J/\psi+X_u | &=& 
a_\psi^\dagger \sum_{X_u} | X_u \rangle \langle X_u | a_\psi = \,
a_\psi^\dagger a_\psi, \\
\sum_{X_c} | X_c \rangle \langle X_c | &=& 1.
\eea
The the usoft Wilson lines only come with the color-octet $J/\psi$ operator, 
${\cal O}^{c\psi}_{\mu}[{\bf 8}^3S_1]$.  
The $\Upsilon$ is a very compact
bound state, due to the large $b$-quark mass. In a multipole expansion, long wavelength gluons interacts 
with the $\Upsilon$ color charge distribution through its color dipole moment since 
the state itself is color neutral. In the theoretical limit of very heavy bottom quark, 
this coupling to the dipole vanishes~\cite{Bobeth:2008prd}. The order of the 
corrections can be estimated by means of the 
``vacuum-saturation approximation''~\cite{Bodwin:1995prd}. A complete set of 
light-hadronic states $\sum_X |X\rangle \langle X|$ 
can be inserted between the $\Upsilon$ operator and the $\jpsi$ operator. 
Notice that the $\Upsilon$ operator is in color-singlet configuration. 
therefore the sum over states is saturated by the QCD vacuum $|0\rangle$ with 
corrections of order $v^4$~\cite{Bodwin:1995prd}. 

Thus we are able to write
\bea\label{SA}
{\cal A}^{abc,a'b'c'}_{\alpha\beta\mu\nu,\alpha'\beta'\mu'\nu'} \,
&\approx& \frac{1}{16\pi^3 \mup} \,
\int \mathrm{d}^4y \,e^{-i\mup/2(1-x)\bn \cdot y} \nn \\
&&\hspace{-4.ex} \times  \,
\langle \Upsilon |\,
 {\cal O}^{\Upsilon\dagger}_\nu[{\bf 1}^3S_1](y) \,
 {\cal O}^{\Upsilon}_{\nu'}[{\bf 1}^3S_1](0)\,
| \Upsilon \rangle \nn \\
&&\hspace{-4.ex} \times \langle 0 |\,
 \tilde{{\cal O}}^{c\psi\dagger}_{\mu}[{\bf 8}^3S_1](y) \,
a^\dagger_\psi a_\psi \,
 \tilde{{\cal O}}^{c'\psi}_{\mu'}[{\bf 8}^3S_1]
|0\rangle \nn \\
&&\hspace{-4.ex} \times  \,
\langle 0 |\,
{\cal J}^{ab\dagger}_{\alpha\beta}(y)
{\cal J}^{a'b'}_{\alpha'\beta'}(0) | 0\rangle  \,.
\eea

Now we substitute the expressions for the $\Upsilon$ and $\jpsi$ shape functions 
as well as the jet function from Section~\ref{FactMat} into Eq.~(\ref{SA}), 
and use 
$p_\psi^2\mathrm{d}p_\psi/(2E_\psi) =  m_b^2 \sqrt{z-4r}\mathrm{d}z/2$.  The 
result for differential decay rate becomes
\bea
\frac{\mathrm{d}\Gamma}{\mathrm{d}z} = \Gamma_0 P[z,r] \,
\sum_\omega |C(\omega,\mu)|^2\,
\int \mathrm{d}k^+ \int \mathrm{d}l^+ J_\omega(k^+)  \,
S_\psi(l^+)  S_\Upsilon(\mup(1-x)-k^+-l^+) \,.
\eea
Thus we obtain the desired factorization theorem in the endpoint region.

\end{document}